\input epsf.sty

\def\be{\begin{equation}}
\def\ee{\end{equation}}
\def\bea{\begin{eqnarray}}
\def\eea{\end{eqnarray}}
\def\bq{\begin{quote}}
\def\eq{\end{quote}}

\def\gappeq{\mathrel{\rlap {\raise.5ex\hbox{$>$}}
{\lower.5ex\hbox{$\sim$}}}}

\def\lappeq{\mathrel{\rlap{\raise.5ex\hbox{$<$}}
{\lower.5ex\hbox{$\sim$}}}}

\documentstyle [12pt]{article}
\begin{document}
\thispagestyle{empty}
\begin{flushright}
{MPI-PhT 96-30} \\
{June 96} \\
\end{flushright}
\vspace{1cm}
\begin{center}
{\large INTERACTING HOT DARK MATTER} \\
\end{center}
\vspace{1cm}
\begin{center}
{Fernando Atrio-Barandela }\\
\vspace{.3cm}
{F{\'\i}sica Te\'orica. Universidad de Salamanca. \\
Plaza de la Merced s/n. 37008 Salamanca, Spain.}
\end{center}
\begin{center}
{Sacha Davidson }\\
\vspace{.3cm}
{Max-Planck-Institut f\"ur Physik\\
F\"ohringer Ring 6, 80805 M\"unchen, Germany}
\end{center}

\begin{abstract}
We discuss the viability of a light particle 
($\sim 30$ eV  neutrino) with
strong self-interactions as a dark matter candidate.
The interaction prevents
the neutrinos from 
free-streaming during the radiation dominated regime
so  galaxy sized density perturbations can survive.
Smaller scale perturbations are damped due to
neutrino diffusion.
  We calculate the power
spectrum in the imperfect  fluid approximation, 
and show that it is damped at
the length scale one would estimate  due to neutrino diffusion.
The strength of the neutrino--neutrino coupling
is only weakly constrained by observations, and
 could be  chosen by fitting the power spectrum
to the observed amplitude of matter density perturbations.
The main shortcoming of our model is that  interacting
neutrinos can not provide the dark matter in dwarf galaxies.
\end{abstract}

\newpage

\section{introduction}

Observations on many scales imply that
most of the matter in the universe is not emitting
electromagnetic radiation \cite{DM}. The ratio  of the
energy density in this unseen
matter component (usually refered as dark matter or DM)
to the critical density today, is defined as $\Omega_o$ and
is under debate.
Various dynamical determinations of $\Omega_o$
on the scales of galaxies and clusters give $\Omega_o \simeq .2 \pm .1$
\cite{DM}.
Observations of large velocity flows suggest
that $\Omega_o$ could be larger than this,
possibly $\Omega_o \simeq 1$ with large  uncertainties \cite{potent, Dekel}.
Theoretical prejudice favours
$\Omega_o = 1$,    because this is more
natural that $\Omega_o < 1$, and because it
is predicted by inflation. 
However, in this case, one needs $\Omega \sim .8$ smooth
on galaxy scales.

Primordial nucleosynthesis implies that the energy density
in baryons satisfies
$.003 < \Omega_B < .06$ \cite{BBN}, which might just be consistent
with the observations of $\Omega_o$. However, it is very difficult to build
models  involving only
baryons that produce the large scale structure and
the Cosmic Microwave Background (CMB) temperature
fluctuations we see today \cite{naoshi,martin},  so it is
usually assumed that the dark matter is made of something other
than baryons.  The list of potential candidates
is long; the popular possibilities are
massive neutrinos, axions, and supersymmetric particles.

In the standard picture, the dark matter (DM)  is 
in thermal equilibrium with the
rest of the Universe at some early epoch,  
but  effectively non-interacting 
when length   scales of interest for structure
formation come into the horizon. In this case, the only free parameter
is the mass of the DM particle.
If it had $m \gappeq 1 $ GeV, it would be non relativistic at
all times of interest for structure formation,
and it would behave like a pressureless fluid.  If it was 
light, it would be relativistic
until cluster scales came within the horizon,
and  perturbations on smaller scales
would be damped by free-streaming. As a result, 
in this second type of model (called hot dark matter or HDM models)
clusters are produced first and must fragment into
galaxies. This ``top--down" galaxy formation scenario is difficult to
reconcile with observations \cite{peebles}
which indicate that galaxies and small scale structure
formed first. 
A model where the dark particles behave non--relativistically
is  therefore prefered. The standard cold dark matter  (CDM) model was found
to have too much power on small scales when normalized to the
amplitude of the measured COBE/DMR temperature anisotropies
and several modifications were introduced. Another 
possibility is a ``warm'' dark matter 
particle with a mass ($\sim$ keV)  intermediate between that of
``hot'' ($\sim$ eV)  and ``cold'' ($>$ GeV) \cite{WDM}. Currently,
models with a mixture of hot and cold dark matter (MDM) are  favoured
 \cite{liddle}.  

It is usually  assumed
that  during structure formation ($T_{\gamma} < 100$ eV) 
the evolution of the Universe
was controlled by electromagnetism and
gravitation, while  new particle
physics takes place at accelerator energies.
However, it is possible that  the dark matter
particles could have some unexpected  interaction
during the epoch of structure formation, despite this 
occuring at a low energy scale. Some
examples of this have been previously studied \cite{RS,CMH,dL,FNAL}.
In this paper, we consider a simple
interacting dark matter model;  we assume that  the dark
matter consists of  light particles
with  a large cross section for scattering
off each other (we discuss later the meaning
of ``large''). For definiteness, we take 
the light particle to be a neutrino with
mass $\sim 20 - 30$ eV (as for HDM).
Our interest is to determine
whether, in the presence of  hot dark matter self-interactions,
density perturbations on scales smaller than
galaxy clusters remain undamped. 
In normal HDM, they are washed out by
free streaming of relativistic particles;
in an interacting model, one would
rather expect the perturbations to 
be damped on a smaller scale by diffusion
(Silk damping).  We will show that this is 
in fact the case, and compute on what
scale they are damped. 

The question of whether ``sticky
neutrinos behave like cold dark matter''
has been previously addressed by Raffelt
and Silk \cite{RS}. They argue that light
($m \sim 20$ eV) interacting
neutrinos will dissipate
small scale perturbations ($<$ galaxy sized)
 by ``Silk damping'', but there
will be no free streaming. 
This result was briefly questioned by
Dicus {\it et al.} in their paper
about neutrino interactions in supernovae \cite{DNPT}, where they
 suggest that cosmological density perturbations in a neutrino
fluid would dissipate
at the speed of sound \footnote{They note, however,
that their flat space results are not directly
applicable to cosmology.}. It 
is well known that  perturbations in a relativistic
perfect fluid in a radiation dominated
Universe neither grow nor are damped \cite{P}. 
However, if the  mean free path of the neutrinos is
non-zero,  they do not constitute a perfect
fluid. There are contributions to the fluid viscosity and
heat conduction coefficients  that are proportional to
the mean free path. The perturbations in the
relativistic neutrino fluid in the early Universe will be damped
by the fluid imperfections, and as Weinberg \cite{W}
showed for Silk damping, the damping length scale  can
be estimated by the  random walk argument used in \cite{RS}. 
Raffelt and Silk's  result is  therefore a correction 
to a perfect fluid analysis, not a contradiction
to it.

The aim of this paper is to  study 
 ``sticky neutrinos'' as dark matter in the
imperfect fluid approximation. Providing that the particle
mean free path is much shorter than
the lengthscale of the density perturbations,
this is a sensible approximation because it
takes into account the non-zero mean free path
of the particles, but avoids the complexity
of the Boltzmann equation.  In section 2, we 
briefly  review the experimental bounds  on
the $\nu \nu$ scattering cross section, and discuss
theories in which such interactions could be strong.
In the first part of the third section,  we
review the
Boltzmann equation and its relation to the fluid approximation,
in the interest of making this paper more self-contained.
In the second part of this section,
we  estimate the scale on which perturbations
in an ``sticky hot dark matter'' Universe would 
be damped. We do this by treating the neutrinos as
a perfect fluid in an expanding
Friedmann-Robertson-Walker Universe to lowest order, and then 
computing the viscosity 
as a perturbation proportional to
the neutrino mean free path.  We briefly discuss the
subsequent behaviour of the sticky neutrinos in
galaxy formation, and calculate
the power spectrum.
In section 4 we show that interacting
neutrinos cannot constitute the halo of dwarf galaxies. 
However, we argue that  this is a potential
problem not only for light fermions but for all dissipationless
DM.  Finally, we present our conclusions.

\section{particle physics}

Despite the precision  to which most Standard
Model quantities are known, there are no experimental
bounds on the $\nu \nu$ or $\nu \bar{\nu}$ scattering
cross section. This is hardly surprising, as it
would be difficult to scatter neutrino beams off
each other with a high enough luminosity  to see
anything within the Standard Model. 
The best constraint on the $\nu \nu$ and $\nu \bar{\nu}$
cross section  comes from the observation
of the neutrinos from SN1987A. These
 had to pass through  the  ``cosmic neutrino
background radiation''  between here and the Large Magellanic Cloud (LMC)
before arriving at the earth.
Approximately the right number of neutrinos were detected, 
which places an upper bound on the $\nu \nu$ interaction.
This constraint has been computed in \cite{KTGR}; we briefly
outline their arguments here.

The ``cosmic  background'' of light stable neutrinos
  has a number density
$n_{\nu} = \frac{7}{8} (T_{\nu}/T_{\gamma})^3 n_{\gamma}$, where
$T_{\nu}/T_{\gamma}$  depends on how many different families of
particles annihilate into photons 
 after the neutrino  gas decouples from the photons.
In this paper, we will assume $ (T_{\nu}/T_{\gamma})^3 = 4/11$,
$n_{\nu} \simeq n_{\gamma}/3$.

The flux of $ E_{\nu} \sim 10$ MeV neutrinos from the supernova will 
arrive at the earth if the mean free path $\lambda$ of a supernova neutrino
is of order the distance $D \sim 2 \times 10^{23} $ cm  from
here to the LMC. The mean free path
is computed in \cite{KTGR} with some care, but can
be roughly estimated as
\be
\lambda^{-1} \simeq \sigma n_{\nu}
\ee
where $\sigma$ is the cross section for a neutrino to
scatter off a $\nu$ or a $\bar{\nu}$.
Requiring $\lambda/ D \gappeq 1$  gives 
\be
\sigma \lappeq 90 ~{\rm GeV}^{-2}, 
~{\rm at~} \sqrt{s_{sn}} = 20 ~ {\rm keV} \label{bd}
\ee
 where $\sqrt{s_{sn}}$
is the centre-of-mass energy of
the collision. Since we assume that the
neutrinos are the dark matter, they
have masses of order 20 ---  30 eV, and $s_{sn} \simeq 2 E_{\nu} m_{\nu} 
\simeq (20$ keV)$^2$.

 We are interested in the neutrino self interaction
cross section at centre-of-mass energy scales near
$ 10$ eV. This is because galaxy scales
come into the horizon when $T \gappeq 10$ eV, and the
neutrinos become non-relativistic shortly thereafter;
if galaxy scale perturbations are going to be damped, it
will be during this period. We therefore need to
scale (\ref{bd}) down to lower values of $s$. 
>From a phenomenological point of view, $\sigma$ can scale
as $\sigma(s) = \sigma(s_{sn})s/s_{sn}$ (due to
exchange of a boson with $m \gg s$), 
or as $\sigma(s) = \sigma(s_{sn})s_{sn}/s$ (exchange of
a boson with $m \ll s$). Galaxy-sized perturbations will be  undamped
in both cases. We will concentrate on a model
where the cross section scales as $s$. The other possibility, that it
would scale as $1/s$, would lead to an interaction strong enough to affect
the dynamics of groups of galaxies today. Long range dark matter interactions
have been considered in \cite{FNAL}.

Sticky neutrino models are not easy to construct. 
The obstacle is that the 
Standard Model neutrinos are members of the same
$SU(2)$ doublet as the electrons, so it is difficult to give
them strong self-interactions without
contradicting present experimental data.
This can be avoided by introducing right-handed gauge
singlet neutrinos \cite{Pec}, but then one has difficulties
with the primordial nucleosynthesis bound on
the number of light degrees of freedom present
at $T \sim 1$ MeV in the early Universe \cite{BBN}:
$N_{\nu}  \lappeq 4$. Attempting to
construct a neutrino mass matrix that is
consistent with this constraint, and
various indications of neutrino oscillations
(solar neutrino problem, atmospheric neutrino
deficit,...)  requires more fine-tuning.

Sticky neutrinos were
originally studied in the context of a modified triplet
majoron model \cite{GR,Pal}, where  the majoron 
(the  nambu-goldstone boson associated with
the spontaneous breaking of lepton number $L$)
was more massive than the
lightest neutrino.    The neutrinos interact
via the exchange of the majoron. However,
the triplet majoron couples to the $Z$, and has not been detected
at LEP, so this model
is ruled out. If one extends the Standard Model
fermion content to include right-handed neutrinos, 
then these can acquire
majorana masses by coupling to
a singlet scalar with a vacuum expectation value, and Dirac masses
with the left-handed neutrinos
by coupling to the Higgs. This is
the singlet majoron model \cite{CMP}, and is
experimentally only weakly constrained.  The angular component
of the scalar is the majoron, and its coupling   
to the light neutrinos is proportional to the
neutrino mass.  This  model can be fine-tuned to produce
interacting  light neutrinos.
The right-handed neutrino majorana masses must be of order
the Dirac masses, to allow a big enough neutrino
scattering cross section by majoron exchange.
To avoid having too many neutrinos at nucleosynthesis,
one can make one family light enough that the right handed
component is not yet in equilibrium,  and  another heavy enough
to decay beforehand. (We have added strong
neutrino self-interactions, so it is possible
to arrange fast decays for a heavy neutrino).
This adds up to three neutrinos
plus the majoron ($= .6 \nu$)  at nucleosynthesis.  One must then impose
flavour symmetries, to prevent the middle
neutrino, who is the dark matter candidate, 
from decaying. This mass matrix can not
explain the solar neutrino problem by  oscillations.

Alternatively, we can disconnect the
exchanged scalar from the neutrino mass generation
mechanism. In this way,
we can get ``three neutrinos at
nucleosynthesis''.  If we introduce a few MeV scalar
with gauge strength couplings to the
dark matter neutrino,  it induces strong enough 
interactions to preserve galaxy scale perturbations, and
may not be present as a relativistic degree of freedom
at nucleosynthesis. We must again arrange the neutrino
mass matrix so that the heaviest neutrino decays before
nucleosynthesis while the middle one, who is our
dark matter candidate, is stable, and the lightest
one's right-handed component is not in thermal
equilibrium at $T \sim $ MeV. The light neutrino
must therefore have a very small Dirac mass,
and very weak coupling to the scalar. There is again no
possibility of explaining the solar neutrino problem.  

We can construct a   neutrino mass
matrix that could explain the solar neutrino
problem if we allow four neutrinos
at nucleosynthesis. We introduce a $\sim$ few  MeV scalar
whose interactions are unrelated to the neutrino
masses,  give it gauge strength couplings to
the 30 eV neutrino, and very weak interactions with
the other two neutrinos, who are lighter. We assume that these other
two neutrinos also have very small Dirac masses, so
their right-handed components are not in
thermal equilibrium at nuclesynthesis.  This makes
four neutrinos at nucleosynthesis. We can then allow
the left-handed components of the two lighter neutrinos to have
a majorana mass difference that explains the solar neutrino
deficit.

Another possibility would be to 
disconnect the interacting hot dark matter
from the neutrinos, and imagine that there is
a light real self-interacting scalar as well as three
``standard'' neutrinos. This would give
3.6 neutrino families at nucleosynthesis,
and the neutrino mass matrix is still free,  but
one loses the great advantage of neutrinos over
other dark matter candidates, which is that they are known to exist. 
In any case,  we
are mainly interested in interacting hot dark matter
from a phenomenological point of view, and do not advocate
any particular model.

\section{Evolution of density perturbations.}

The aim of this section is to discuss the formalism 
for calculating the evolution of linear perturbations in
interacting dark matter, and apply it to the case of
``sticky neutrinos''. In the first part, we  review the Boltzmann
equation and its relationship to the  fluid
approximation. We  discuss how to compute the
viscosity $\eta$, and the heat conduction $\chi$ for a slightly imperfect
fluid, and  determine their contribution to
the damping of perturbations. In the second part, 
we calculate $\eta$ for a neutrino fluid,
and show that the  estimate of the
damping scale made by Raffelt and Silk \cite{RS} is  essentially correct.

We use the metric
\be
ds^2 = dt^2 - a^2(t) [\delta_{ij} + h_{ij}(\vec{x},t)]dx^i dx^j ~.
\label{AAA}
\ee
where $h_{ij}$ is the metric perturbation in
synchronous gauge. 
We are interested in the effect of particle interactions
on the evolution of matter density perturbations within
the horizon. On those scales, metric
perturbations are negligible compared to pressure
gradients, damping and other hydrodynamical effects \cite{sasaki, E}.
We therefore  neglect the metric
perturbations in  our analytic estimates,
although they are included in our numerical
calculation of the power spectrum.

We define $k$ to be the magnitude  of the comoving
three-momentum, and $\gamma_i$ its directional
cosine: if $|p_i p^i|^{1/2} = p$ is the magnitude of the physical
momentum, $p = k/a$,    $p_i =-  k \gamma_i $, 
and $p^i = a^{-1} p \gamma_i$.  Roman indices run from 1 to 3, and
greek from 0 to 3. 

\subsection{The Boltzmann equation and the fluid approximation.}
The Boltzmann equation describes the evolution of the phase space 
density of the particles   $f(x^i,p_i,t)$,
and provides a framework in which one can calculate
the growth and survival of density perturbations  once
they come within the horizon.  It can be written as
\be
{\cal L}[f]  = C[f] \label{be}
\ee
where ${\cal L}[f]$ is the derivative of 
$f$ along the  path in phase space of
the particles, and $C[f]$ is
the collision integral.  If the
particles described by $f$ have no collisions,   $f$ is
constant along a particle's path
in phase space, or ${\cal L}[f] = 0$.   In a relativistic fluid under
the influence of gravitation, ${\cal L} [f]$ is defined as \cite{K+T}
\be
{\cal L} [f] = p^{\mu} \frac{\partial f}{\partial x^{\mu}} -  
\Gamma^{\mu}_{\nu \beta} p^{\nu} p^{\beta} 
 \frac{\partial f}{\partial p^{\mu}} ~.
\label{Lf}
\ee
$C[f]$ parametrizes the changes in the distribution function due to
the particle collisions. The  integral representing the
change in the distribution function $f_a$ for particle $a$
due to the process $a + b \rightarrow c + d$ is \cite{K+T}
\be
C[f_a] =\frac{1}{2}  \int d\Pi_b d \Pi_c d \Pi_d 
\hat{\delta}^4
|{\cal M}_{ ab \rightarrow cd}|^2 (f_a f_b - f_c f_d) \label{Cf}
\ee
where 
$|{\cal M}|^2$ is the matrix element squared for
the process $a + b \rightarrow c + d$, including spin
sums and averages, and   
\begin{eqnarray}
d\Pi &=& \frac{d^3p}{2 E (2 \pi)^3}\\
\hat{\delta}^4 &=& (2 \pi)^4 \delta^4( p_a + p_b - p_c - p_d)~.
\end{eqnarray}
Implicitly, in equation (\ref{Cf}) we have used 
Maxwell-Boltzmann statistics by neglecting the occupation
in the final phase space. 
This approximation is justified if the chemical
potentials of all particles are negligible at the
temperature scales involved.
But even in this limit, the solution of the Boltzmann equation is 
 complicated.

One can avoid dealing with the
collision term by integrating (\ref{be}) over
quantities that are conserved in the interaction
decribed by $C[f]$ \cite{K}. This eliminates 
the right-hand-side of (\ref{be}). For instance,  if
$p_{\mu}$ is the four-momentum of
the particle $a$, one can write
\be
\int d \Pi_a p^{\mu}_a {\cal L}[f_a] = \int d \Pi_a p^{\mu}_a C[f_a]
\ee
and it is easy to show from symmetry considerations that
the integrand on the right hand side  is proportional to
$p^{\mu}_a +  p^{\mu}_b -  p^{\mu}_c - p^{\mu}_d$, so 
the integral is zero. Therefore, one obtains
\be
\int d \Pi_a p^{\mu}_a {\cal L}[f_a] = 0 ~. \label{ce1}
\ee
In the case of sticky
neutrinos, the scattering interaction conserves
particle number, so we also have 
\be
\int d \Pi_a  {\cal L}[f_a] = 0 ~.\label{ce2}
\ee
Defining the stress-energy tensor for a single  species as
\be
T^{\mu \nu} = 2 \int d \Pi f p^{\mu} p^{\nu}~, \label{Tmunu}
\ee
and the number density vector
\be
N^{\mu} = 2 \int d \Pi f p^{\mu}~, \label{N}
\ee
it is easy to check,  in a Friedmann-Robertson-Walker
Universe, that (\ref{ce1}) and (\ref{ce2})
are equivalent to the usual  conservation equations
\be
T^{\mu \nu}_{~~; \nu}= T^{\mu \nu}_{~~, \nu} + 
 \Gamma^{\mu}_{ \nu \kappa} T^{\nu \kappa} 
+ \Gamma^{\nu}_{ \nu \kappa} T^{\mu \kappa}   = 0 \label{Tcons}
\ee
and 
\be
N^{\mu}_{~ ; \mu} = 0~.
\ee
It is clear, however, that since
these are  only conservation equations,
they do not contain enough information to
specify the dynamics. We still need to 
solve the Boltzmann equation (\ref{be}) to obtain $f$.
Since the Universe was homogeneous and
isotropic to better than one part in $10^5$
when it was radiation dominated, 
one can solve the Boltzmann equation for
perturbations  away from the homogenous
and isotropic  background \cite{P}.
A  second possibility,
which can only be used for interacting 
particles, is to
perturb away from a perfect fluid. 
We shall use this second approach for sticky
neutrinos, since we assume they
are strongly interacting.  
These two schemes are not
identical,  because in the first case
the background is  a homogeneous and isotropic equilibrium
distribution $f \sim \exp \{ - E/T(t)\} $, and 
the perturbation in $f$ corresponds to a density perturbation.
In
the second case,  one perturbs  away from a locally
thermal distribution $f \sim \exp \{ - E/T(x,t) \} $,
so density perturbations are already present in the background
solution.
It is a two step procedure to find the evolution of
the density perturbations in the fluid approximation. We
first estimate the magnitude of the perturbation away from
local thermal equilibrium using the Boltzmann equation.
Then we calculate the contributions of these
perturbations to the stress-energy tensor, and
solve for the evolution of the density
perturbations from the conservation equation (\ref{Tcons}). 
The perturbations away from local thermal equilibrium
introduce new terms in
(\ref{Tcons}) which can affect the growth of the density perturbations.

A perfect fluid is locally in  thermal
equilibrium, so we can write the phase space density for
particles consituting an imperfect fluid as
\be
f = g^{(0)}(E,x^i,t) + g^{(1)} (E, p_i, x^i, t)
\ee
where
\be
 g^{(0)}(E,x^i,t) = \frac{g}{(2 \pi)^3} \exp \{ - [E  - \mu(x,t)]/T(x,t) \}
\ee
is the local equilibrium distribution, $g$ is
the number of degrees of freedom of the particle, and
$E = U^{\alpha} p_{\alpha}$, where $U^{\alpha}$ is
the four-velocity of a particular observer.
$g^{(1)}$ measures the departure of the distribution
from exact local equilibrium.

If the particles involved in an interaction are
in local thermal equilibrium, then the collision
term in the Boltzmann equation is zero; however,
$g^{(0)}$ does not neccessarily satisfy the left
hand side of the Boltzmann equation. Assuming that
$g^{(1)} \ll g^{(0)}$, we 
have 
\bea
 E \frac{\partial g_a^{(0)}}{\partial t} +
\frac{p \gamma_i}{a} \frac{\partial g_a^{(0)}}{\partial x^i} 
+ H p^2 \frac{\partial g_a^{(0)}}{\partial E} = 
 \frac{1}{2} \int d\Pi_b d \Pi_c d \Pi_d \hat{\delta}^4
|{\cal M}_{ ab \rightarrow cd}|^2  \\
\times (g_a^{(1)} g_b^{(0)} + 
g_a^{(0)} g_b^{(1)} - g_c^{(1)} g_d^{(0)} - 
g_c^{(0)} g_d^{(1)}) ~. \label{be3}
\eea
In principle, to correctly compute the deviations
from perfect fluid behaviour, we still have to 
evaluate the collision integral. However, if we 
only wish to estimate viscosity and heat conduction
coefficients, the right hand side can be taken \cite{K}
(to within a factor of $\sim$ 4) as its first term
$ \sim  E_a \Gamma g^{(1)}$, where $\Gamma$ is the 
thermally averaged interaction
rate for the reaction described by $C[f]$. This gives
\be
g^{(1)} \simeq \frac{ {\cal L}[g^{(0)}]}{ E_a \Gamma} ~. \label{g1}
\ee
To get the damping scale of the perturbations
correctly, we would need a better approximation,
as can be found, for instance, in \cite{CE}.
A more accurate approach would require a numerical
integration of the coupled Einstein and Boltzmann equations
that describe the behaviour of metric perturbations and 
the density field of the different particles \cite{MAB}.

If  the mean time between particle
collisions is $\tau = \Gamma^{-1}$, and $t$ 
is the timescale associated with the density
perturbation in the fluid, (\ref{g1}) implies
$g^{(1)} \sim g^{(0)} \tau/t$. So 
 $g^{(1)} \ll g^{(0)}$, and
the fluid approximation is consistent, 
provided that the size of
the density perturbations in the early Universe
is much longer than the mean free path of
the particles.  In this limit, the phase
space density to first order in $\tau$ is
\be
f = g^{(0)} + \frac{\tau}{E}  {\cal L}[g^{(0)}] \label{f}
\ee

We have not yet defined $\mu$ or $T$, or
discussed in which reference frame we are measuring
the energy $E$.  In principle $E = p^{\alpha} U_{\alpha}$,
so we must specify $U_{\alpha}$.
We follow Landau and Lifschitz \cite{LL}, and 
take $U_{\alpha}$ to be the velocity associated
with energy flux. (Note that
this is not the choice made in \cite{W},
so we can not immediately compare to
Weinberg's equations). This is natural in the approximation
scheme we have used,  because
\be
T^{0 i} = 2 \int  d \Pi E p^{i} g^{(1)} = 2 \int d \Pi  p^{i} {\cal L} 
[g^{(0)}] = O(\tau^2) \simeq 0 \label{Toi}
\ee
where the last equality follows from 
(\ref{ce1}). We define $T$  such that the equilibrium expression
for the energy density is equal to the  energy density
in the reference frame where there is no energy
flux, and $\mu$ such that the equilibrium expression
for the number density is equal to the
measured number density in the same reference frame.
(In other words, the perturbations $g^{(1)}$ will
not contribute to $\rho$ or $n$ in the reference
frame where $T^{0i} = 0$). 

The general form of the stress-energy tensor and the particle
flux vector $N^{\mu}$
for a slightly imperfect relativistic fluid can be determined 
by requiring the entropy to be non-decreasing \cite{W,LL}.
In the rest frame of $U$, the velocity associated with energy flux,
they are \cite{LL}
\be
a^2 T^{\mu \nu} = \left[
\begin{array}{cccc}
a^2 \rho & 0 & 0 & 0 \\
0  & 
P - (\zeta - \frac{2}{3} \eta) \partial_{\alpha}U^{\alpha} & 
- \eta ( \partial^1 U^2 + \partial^2 U^1) &
- \eta ( \partial^1 U^3 + \partial^3 U^1) \\
0 &- \eta ( \partial^1 U^2 + \partial^2 U^1)  
& P - (\zeta - \frac{2}{3} \eta) \partial_{\alpha}U^{\alpha} & 
- \eta ( \partial^3 U^2 + \partial^2 U^3) \\
0 & - \eta ( \partial^3 U^1 + \partial^1 U^3)
&- \eta ( \partial^3 U^2 + \partial^2 U^3)  
& P - (\zeta - \frac{2}{3} \eta) \partial_{\alpha}U^{\alpha}  
\end{array} \right]
\label{tensor}
\ee
and \be
N^{\mu} = (n,   ~\frac{\chi}{a} \left( \frac{n T}{\rho +P} 
\right)^2  \partial^i \left(
\frac{\mu}{T} \right) )~.  \label{Nmu}
\ee
Formal expressions for the heat conduction coefficient
$\chi$ and the bulk and ordinary viscosities $\zeta$ 
and $\eta$ have been elegantly calculated in \cite{I+V} for a generic
relativistic fluid. However, their integrals are
difficult to evaluate. 
We can  simply approximate the fluid parameters
 by substituting (\ref{f}) into (\ref{Tmunu})
and (\ref{N}),
and comparing to the formulae for $T^{\mu \nu}$ and
$N^{\mu}$. To extract $\chi$ from
$N^{\mu}$ one must  use the constraint (\ref{Toi}).
We will do this for sticky neutrinos in the next subsection.

Once $T^{\mu \nu}$ for an imperfect fluid has been calculated, 
the conservation equation (\ref{Tcons}) can be used to study
the evolution of density perturbations on scales much
longer than the particle interaction length.  As noted
in \cite{W}, perturbations in a fluid of relativistic
particles are principally damped by viscosity.
If we neglect the   metric perturbations
(reasonable for sub-horizon sized modes) and the heat conduction
coefficient, and  decompose the density perturbation in
Fourier modes,  (\ref{Tcons})  implies
\be
\dot{\delta} = -\frac{4}{3}\theta  
\label{flatdelta}
\ee
\be
 \dot{\theta} + \frac{\dot{a}}{a} \theta
 =  p^2 (\frac{\delta}{4} -  \frac{ \eta}{\rho} \theta)
\label{flattheta}
\ee
where  $\theta \equiv i k_j  U^j$,
$\delta \equiv \delta \rho/ \rho$, $k$ is the comoving three-momentum
and $p = k/a$ is the physical three momentum. These are the equations
for the longitudinal component ( $\vec{U} \parallel \vec{p})$
of the perturbation; the transverse component is rapidly damped \cite{W}.
One can see from (\ref{flatdelta}) and (\ref{flattheta}) that
the perturbation oscillates and damps at a rate 
\be
\Gamma \sim \frac{2p^2 \eta}{3 (\rho + P)}
\ee
due to the fluid viscosity $\eta$. Including heat conduction $\chi$ 
and bulk viscosity $\zeta$ gives
\cite{W} 
\be
\Gamma = \frac{p^2}{2(\rho + P)} \left[ \zeta + \frac{4}{3} \eta
+ A \chi \right] \label{G}
\ee
where $A$ is a
function of various fluid parameters
\be
A =\left( \frac{ \partial \rho}{\partial T} \right)_n^{-1}
\left[ \rho + P - 2T  \left( \frac{ \partial P}{\partial T} \right)_n
+ v_s^2 T \left( \frac{ \partial \rho}{\partial T} \right)_n
- \frac{n}{v_s^2} \left( \frac{ \partial P}{\partial n} \right)_T \right] ~~.
\ee
 $v_s$ denotes the speed of sound. This function $A$ goes to
zero as particles become extremely relativistic.

\subsection{Perturbation Damping for Sticky Neutrinos}

The length scale on which neutrino diffusion
damps density perturbations increases with
the age of the Universe until the neutrinos
become non-relativistic. We can estimate the comoving
scale on which density perturbations will be
damped by treating the neutrinos as relativistic
until $T \simeq m_{\nu}/3$, and neglecting any further motion
after this temperature. A similiar approximation provides
the ordinary neutrino free streaming scale to within
a factor of three, and since we only know
the fluid viscosity and heat conduction to within
the same order, it should be adequate.

In the extreme relativistic limit,
the bulk viscosity $\zeta$ is zero, 
so we will neglect it.
To compute the viscosity $\eta$, we note
from \cite{LL} that in the reference frame where 
$U = (1, 0, 0, 0)$ 
\be
T^{12} = - \eta \left( \frac{ \partial U^{1}}{\partial x_2} +
\frac{ \partial U^{2}}{\partial x_1} \right) ~.
\ee
 Substituting (\ref{f}) into
(\ref{Tmunu}), we find 
\be
T^{12} = \frac{ \tau}{2 a^2} \int p^2 dp d \Omega \gamma^1 \gamma^2 
p^{\mu} \frac{ \partial U^{\alpha}}{\partial x^{\mu}}
\frac{ \partial g^{(0)}}{\partial U^{\alpha}} + ...
\ee
where the ``...'' indicate the terms
which are zero when the angular integral is performed.
With the identity
\be
\int d \Omega \gamma^i \gamma^j \gamma^k \gamma^l = 
\frac{4 \pi}{15} (\delta^{ij} \delta^{kl}
  +\delta^{ik} \delta^{jl} +\delta^{il} \delta^{kj})
\ee
this gives in the relativistic limit
\be
\eta = \frac{ 4 \tau n T}{5} \label{eta}
\ee
which agrees with the result in \cite{W,Mis,T}. 

The heat conduction coefficient $\chi$ can be
estimated by substituting (\ref{f}) into (\ref{N}),
(with  the constraint (\ref{Toi})) and comparing
to (\ref{Nmu}).
In the relativistic limit, we get
\be
\chi = \frac{4 \tau n}{3}
\ee
which is a factor of 3 smaller than the result
in \cite{W,T} but is  within the errors of our approximation. 

We can roughly estimate the physical length
scale $\ell_D$ on which perturbations will be damped  as
$\ell_D \simeq 2 \pi/p_D$, where $p_D(t)$ is 
is determined from (\ref{G}) by setting 
\be 
\int ^{t_{nr}} \Gamma(t) dt = 1
\ee
and using the relativistic expression
for $\eta$ (\ref{eta}). We neglect heat conduction
since $A \simeq 0$ in the relativistic limit,
and perturbations are principally
damped by viscosity. 
 $t_{nr}$ denotes the moment when neutrinos become non--relativistic
and occurs when $T \simeq  m_{\nu}/3 \simeq 10$ eV. 
The  physical lengthscale  on which perturbations
are damped is  
\be
\ell_D \simeq  2 \pi \sqrt{ \frac{ \tau t_{nr}}{15}}~~.    \label{lD}
\ee
This is approximately the distance one
would estimate that a relativistic neutrino
could travel via a random walk in time $t$:
a neutrino can take $N = t/\tau$ steps
of length $\tau$, so on average each neutrino
travels a distance $\sqrt{N} \tau$.  One expects
the neutrino number density to become
homogeneous on scales shorter than $\sim t/\sqrt{N}$, 
which was  the estimated damping scale
of neutrino density perturbations  given in  \cite{RS}. To summarize,
if one includes interactions
for a hot dark matter candidate, perturbations
are damped by  diffusion rather than by free-streaming.
The length scale on which perturbations are washed
out is therefore shorter, and the power spectrum
should look more like that of cold dark matter,
as claimed in \cite{RS}.

We have neglected mass effects in (\ref{lD}), because
our fluid parameters are only correct to within factors of 3 or 4.
To properly determine the damping scale, we would need
accurate determinations of $\chi$, $\eta$ and
$\zeta$, valid as the temperature drops past the 
neutrino mass.  We could then integrate $\Gamma$ over
all times to determine $\ell_D$. 
Instead,  to check that
our relativistic approximation is acceptable, we have 
calculated $\chi$ and $\eta$ in the non-relativistic
limit, matched them to the relativistic versions at
$T=m_{\nu}$, and calculated $\ell_D$. 
The damping scale is (to
within a factor of 2) the same as in (\ref{lD}).

The comoving scale  up to
which density perturbations are washed out by
neutrino diffusion is approximately
\be
\lambda_D = \frac{2 \pi}{k_D} 
= \frac{T_{nr}}{T_0} \ell_D \simeq  3 
  \sqrt{\frac{\tau}{{\rm Mpc}}}~ Mpc 
\label{lDco}
\ee
where $\tau$ is the co-moving mean free path  at $T_{nr} = 10$ eV.
 As discussed in section 2, the
neutrino mean free path is bounded below by the arrival
of the neutrinos from SN1987A. This means that the 
scale on which perturbations are damped is also
bounded below, and we need to check that galaxy-sized perturbations
do indeed  survive.

The supernova bound implies that
the timescale for neutrino self-interactions 
at $t_{nr}$ (when $T_{\nu} = 10$ eV)
must satisfy
\be
\tau_{nr} \gappeq \frac{n_0 ~\sigma_{sn}}{n_{nr}\; \sigma_{nr}} \tau_{sn}
= \left( \frac{T_0}{T_{nr}} \right)^3 \left( \frac{s_{sn}}{s_{nr}} 
\right)^{\pm 1} \tau_{sn}
\ee
where a 0 subscript means the present value of a 
given magnitude, $n$ is the neutrino
number density, $\sigma$ is the cross section, and $\tau_{sn}$ is
the distance to the LMC. $s_{nr}$ and $ s_{sn}$ 
are respectively  the centre-of-mass energy for the
neutrino self--interaction at $T_{\nu} = 10 $ eV,
and for scattering a supernova neutrino off the neutrino CMB. 
The  ratio  $s_{sn}/s_{nr}$
is to the power $+1$ if the cross section scales as $s$
(exchanged boson mass $> \sqrt{s_{sn}}$),
and to the power $-1$ for a cross section scaling
as $1/s$ (exchanged boson mass $< \sqrt{s_{nr}}$).
For a non-renormalisable
cross section, the supernova bound  
implies that perturbations on comoving scales
$\lappeq 10^{-2}$ Mpc
are damped by  neutrino  diffusion  (or $  \sim 10^{-7}$ Mpc
for a cross section that scales as $1/s$).
The supernova bound on the
cross section  is therefore consistent with  the survival
of galaxy scale perturbations, whether the neutrino
stickiness is due to the exchange of light or heavy
particles.

\subsection{Evolution in the non--relativistic regime and final power spectrum.}

We have seen that the sticky neutrinos behave quite differently from
standard neutrinos while they are relativistic. However, once they
are non-relativistic, they become a  pressureless
dust to lowest order, like most other dark matter candidates. 
If the neutrino cross section scales as $s$, then the interactions
become weaker as the temperature drops, and our fluid
approximation breaks down. However, if the neutrino interaction
timescale is of order the horizon size
then the interaction is irrelevant. On the other
hand, if our fluid approximation is still valid, 
then we know that the viscosity and heat conduction
coefficients are proportional to the pressure \cite{K},
and the pressure is negligible. So we expect 
the non-relativistic, interacting neutrinos to behave like standard 
neutrinos, if the interaction is short range. A longer range
interaction (exchanged boson mass $\ll m_{\nu}$)
is potentially more complicated, particularily if
the interactions are still strong today.
Dark matter with long-range interactions
has been considered in \cite{FNAL};
we neglect it here. 
Constraints on dark matter self-interactions from
galaxy formation and evolution in
the non-linear regime have been estimated in \cite{dL}.
The authors argue that the average interaction length of dark
matter today must be longer than 100 Mpc. Sticky neutrinos
who interact via the exchange of a $m \sim $ MeV boson
and who are  consistent with the supernova bound
are essentially non-interacting today (the interaction
rate scales as $a^{-5}$ or $a^{-7}$), so  they easily satisfy the
constraints listed in \cite{dL}.

Based on our analytic results,
we expect the power spectrum of interacting 
HDM to  be  similar to
standard CDM on large and intermediate scales,
but damped on small scales by neutrino diffusion.
We are working in the fluid approximation,
so to compute the power spectrum, we must numerically
integrate the  Einstein equations and the energy 
conservation equation (\ref{Tcons}) for an
imperfect fluid. This is considerably simpler than integrating 
the coupled Einstein and Boltzmann equations. 

Let us briefly outline here the equations and
approximations used.
We perform our calculations
in the synchronous gauge using the metric (\ref{AAA}).  The equations
governing the evolution of  perturbations are
more easily written in Fourier space. If we only
consider only scalar--type  perturbations, the metric perturbations
can be written in terms of two scalar fields $h(k,t)$
and $\eta( k,t)$ as follows \cite{MAB}:
\be
h_{ij}(x,t) = 
 \int d^3k e^{i\bf kx}[k_ik_jh(k,t) +\{k_ik_j -{1\over 3}\delta_{ij}
  6\eta(k,t)\} ]~.
\ee
As in the FRW case, the equations governing
the evolution of matter density perturbations can be 
derived from the conservation equation (\ref{Tcons}).
For a relativistic fluid, with viscosity but no
bulk viscosity or heat conduction,
equations (\ref{flatdelta}) and (\ref{flattheta})  become:
\be
\dot{\delta} = - \frac{4}{3}\left(\theta  -{1\over 2} \dot h \right)
\label{curvedelta}
\ee
\be
 \dot{\theta} + {\dot a\over a}\theta = p^2 \left(\frac{\delta}{4} 
-  \frac{ \eta}{\rho} (\theta- 3\dot\eta) \right)
\label{curvetheta}
\ee
This results differs 
from \cite{peebles}   or \cite{MAB}
because  our matter does not behave as a perfect fluid.
As the temperature drops past the neutrino mass,
the neutrinos no longer behave as relativistic particles.
The evolution of density perturbations in the $E \sim m$
regime is complicated by the time-dependent  relation 
between the neutrino momentum and energy \cite{MAB}.
This adds terms to (\ref{curvedelta}) and (\ref{curvetheta})
and makes the numerical calculation of the pressure and
density  more difficult \cite{MA}. 
To simplify the calculation, we assume
the transition from a relativistic gas to presureless dust
to be instantaneous. We also assume that neutrino interactions
are frozen in the non-relativistic regime due to the small
mean velocity. As a result, we treat the neutrino gas
as CDM particles with zero velocity in the synchronous
gauge, whose evolution is given by:
\be
\dot{\delta} = {1\over 2} \dot h 
\label{delta_no_viscosity}
\ee

As previously noted, perturbations
in a relativistic fluid are damped by viscosity. 
The off-diagonal spatial elements of the  stress-energy
tensor (parametrized by the viscosity)
are also gauge-invariant, which makes them simple
to calculate. We therefore include only the 
viscosity (no heat conduction
coefficient and no bulk viscosity)
in our numerical calculation of the power spectrum.
We take the physical interaction
time $\tau$ to scale as $a^5$ 
(comoving $\tau \sim a^4$), and
 assume an initial  Harrison--Zel'dovich
spectrum of perturbations from inflation. 

In Fig.1 we give the numerical integration 
of the previous set of differential equations
coupled with the equations for the evolution 
of metric perturbations, baryons and photons.
We used a modified version of the COSMIC package made publicly
available by Berstchinger and Ma, and described in \cite{MAB}.
The dashed line corresponds to standard CDM 
and  is plotted  for comparison. The three solid
lines correspond, in decreasing amplitude,
 to zero viscosity (zero mean free path) and
comoving mean free paths at $T_{\gamma} = 10$ eV
of $ 10^{-2}$ Mpc and
$ 0.1$ Mpc. In the
case of zero viscosity, the neutrino perturbations
oscillate as acoustic waves around the CDM power spectrum. The
behaviour is similar to the baryon-photon plasma
where the interactions cause the perturbations
in the baryon component to oscillate
as sound waves.   When the viscosity is introduced
the oscillations are damped approximately at the scale given
by the diffusion length.
The effect of viscosity is to damp 
the power spectrum at the scale given by
(\ref{lDco}): 3 and 7 Mpc${-1}$ respectively
for co-moving mean free paths of $0.01$ and
$0.1$ Mpc. 
The damping is a purely fluid effect 
and it has nothing to do with gravitation
as was already remarked in \cite{W}.

One should notice that the overall power spectrum
amplitude is not completely accurate on scales 
$k\simeq 0.3$Mpc$^{-1}$. We assumed, when calculating the
power spectrum, that the neutrino gas 
underwent an instantaneous
transition from radiation to pressureless dust
at $T_{\gamma} = 10$ eV.
For the  purpose of showing the effect
of damping on small scales due to viscosity,
the instantaneous transition approximation is quite accurate,
and  simplifies the calculations.

To summarize, neutrino viscosity helps to reduce
power on small scales while keeping most 
of the features of CDM. In this way, 
it alleviates the problems that plague the standard CDM
model, such as large scale pairwise velocity 
dispersion, that come from having too little power
on large ($\ge$ 30 $h^{-1}$ Mpc) scales relative to
small ($\le$ 10 $h^{-1}$ Mpc) ones.
A detailed analysis of the success of our model
will require N--body simulations to compare with observations
as performed, for example, by  \cite{WDM}.
However, such analysis goes beyond the scope of the present paper.
\begin{figure}[htb]
\begin{center}
\epsfxsize=12cm \epsfysize=10cm \epsfbox{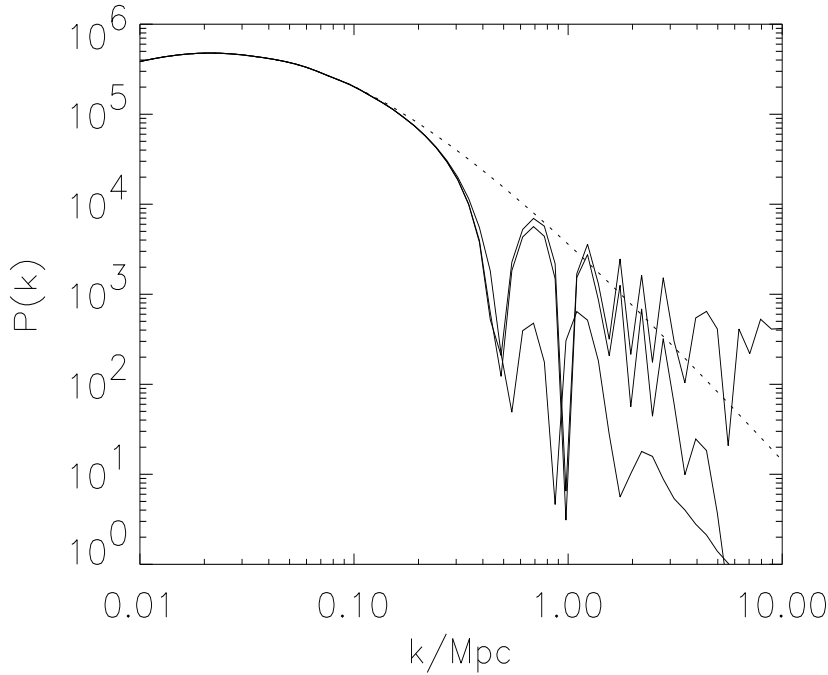}
\caption{ Power spectrum for three different 
``sticky neutrino'' models. The solid lines
correspond (in decreasing amplitude) 
to interactions with no neutrino diffusion, and 
with co-moving mean free paths at
$T_{\gamma} = 10$ eV of .01 Mpc and  .1 Mpc. These
give rise to estimated  damping scales of 7 Mpc$^{-1}$
and  3 Mpc$^{-1}$, respectively. The dashed line
correspond to standard CDM and is plotted for
comparison. The y-axis scale is arbitrary.
We took $H_o = 50$Km s${-1}$Mpc${-1}$.}
\end{center}
\end{figure}

\section{Phase Space Constraints}

The main argument against neutrinos as DM candidates
was put forward by Gunn and Tremaine \cite{GT}. They argued
that the large phase space density
of galaxies was close to the density of a degenerate
Fermi gas and they virtually ruled out neutrinos as dark matter
candidates.
The argument has been refined since \cite{madsen1, madsen2}.
Currently, the tightest constraints 
come from two dwarf spheroidal galaxies within our local group:
Draco and Ursa Minor.
In those galaxies, if the spatial distribution
of dark matter and luminous stars is similar, the central dark
matter densities needed to explain the kinematic data
are $\rho_{DM} = 0.5 - 1M_\odot pc^3$.  For non--interacting 
neutrinos, the conservation
of the phase space density implies \cite{gerhard}:
\begin{equation}
m_\nu \ge 170 eV \left({1kpc\over r_{DM}}\right)^{3/4}
        \left({0.1 M_\odot pc^{-3}\over\rho_{DM}}\right)^{1/8} g_\nu^{-1/4}
\end{equation}
To overcome this limit and bring $m_\nu$ down to 
$30 $ eV, the core radius of the 
DM, $r_{DM}$, should be almost two orders of magnitude larger than the one
of the luminous stars, making the mass to light ratio of the system
close to $10^6 M_\odot/L_\odot$. For sticky neutrinos the 
bound is very similar \cite{RS}. 

However,
the halo of dwarf galaxies is not only a difficult test for fermionic 
DM, but any dissipationless galaxy formation scenario.
The density of an object that formed without dissipation uniquely specifies
its redshift of formation \cite{lake}:
\begin{equation}
1+z_{turn} \simeq 4 h_{50}^{-2/3}
        \left({\rho_{D}\over 10^{-2}M_\odot pc^{-3}}\right)
\end{equation}
where $\rho_{D}$ is the central density of the system.
For Draco and Ursa Minor, the previous equation shows that
$(1+z_{turn}) \simeq 30$. In  CDM,
MDM  and similar scenarios \cite{lucchin} the first structures
turn around at about $(1+z_{turn}) \simeq 10$ and to bring down the previous
limit would be necessary to have a DM core radius 4 times larger than
the luminous matter and $M/L \simeq 1000 M_\odot/L_\odot$ for these
dwarfs, which is also rather high.
Lacking a clear understanding of star and galaxy formation one can not disregard
the possibility that a significant fraction of the halo is baryonic 
or dissipational \cite{Evrard}, which would
weaken the limit on the neutrino mass.

\section{Discussion}

As mention in the introduction, the ``standard'' CDM
model, when normalized to the amplitude of temperature
anisotropies measured by COBE/DMR, has too much power
on small scales.  Several modifications
were proposed that agree better with observations,
such as, for instance,
hot plus cold dark matter models, CDM models with cosmological
constant, tilted power spectrum and variants thereof. 
However, if one is going to add free parameters to
one's dark matter model, it is in principle
as reasonable to add interactions for a single component
as to mix two (or three) components. As an example of
this, we have considered a dark matter
model consisting of  $\sim$  30 eV neutrino
with a large $\nu \nu$ scattering cross section.
We have argued in this paper that these
interacting neutrinos,
although not easy to produce in a reasonable
particle physics model, may be  a viable dark matter candidate. 
Unlike ordinary hot dark matter, they do not free-stream. 
In section 3 we estimated
the damping scale associated with neutrino diffusion
and calculated the power spectrum.
The current upper limits on the $\nu \nu$ cross section
permit a damping scale  shorter than $1 Mpc$ 
and accordingly galaxy scale density perturbations will
survive. Stronger upper limits on the cross section
will increase the damping scale and could invalidate the model.
The main positive feature of our model is that
the power on  small scales is reduced by neutrino
diffusion while on larger scales we retain the main feautures of 
standard CDM. The neutrino--neutrino cross section,
which is the  only  free parameter, could possibly 
be tuned to give ``the right
amount of power on small scales''.
The main handicap  is that
interacting neutrinos, like ordinary neutrinos, 
can not constitute the halos of
dwarf galaxies. However, it can be argued that
dwarfs are difficult to explain in any dark matter scenario,
so this may not be a fatal problem.

\subsection*{acknowledgements}
We would like to thank Georg Raffelt for useful
conversations, and F. A.-B. is grateful to the
 Sonderforschungsbereich 375-95
for support  when this work was completed.


\begin{thebibliography}{222222}

\bibitem{DM} T. Padmanabhan.``Structure Formation in the Universe",
(1993), Cambridge University Press, Cambridge (England).
\bibitem{potent} A. Dekel, E. Bertschinger, S.M. Faber,
{\it Astrophys. J.} {\bf 364} (1990) 349. 
\bibitem{Dekel} A. Dekel, {\it Ann. Rev Astron. Astrophys.} 
{\bf 32} (1994) 371.
\bibitem{BBN} see, for instance, B.D. Fields, K. Kainulainen, K.A. Olive,
D. Thomas, {\it Model Independent Predictions of Big Bang
Nucleosynthesis from $^4$He and $^7$Li: Consistency
and Implications}, astro-ph/9603009; C. J. Copi, D. N. Schramm, 
M. S. Turner, astro-ph/9606059.
\bibitem{naoshi} N. Sugiyama, N. Gouda, {\it Progress of Theoretical 
Physics} {\bf 88} (1992) 803. 
\bibitem{martin}M. White, D. Scott, J. Silk,
{\it Ann. Rev. Astron. Astrophys.} {\bf 32} (1994) 319.
\bibitem{peebles} P.J.E. Peebles. ``Principles of Physical Cosmology"
(1993), Princeton University Press, Princeton (New Jersey).
\bibitem{liddle} A.R. Liddle, D.H. Lyth, R.K. Schaefer, Q. Shafi, P.T.P. Viana
{\it Pursuing parameters for critical--density 
dark matter models}, astro-ph/9511057.
\bibitem{WDM} S. Colombi, S. Dodelson, L.M. Widrow {\it Ap. J.} {\bf 458} (1996) 1.
\bibitem{RS} G.G. Raffelt, J. Silk, {\it Phys. Lett.} {\bf B192} (1987)
 65.
\bibitem{CMH} E.D.  Carlson, M. Machacek , L.J.  Hall, {\it Ap. J..} 
{\bf 398 } (1992) 43;
M. Machacek , {\it Ap. J.} {\bf 431} (1994) 41.
\bibitem{dL} A.A. de Laix, R.J. Scherrer,
R.K. Schaefer, {\it Ap. J.} {\bf 452} (1995) 495.
\bibitem{FNAL} B. Gradwohl, J.A. Frieman, {\it Ap. J.} {\bf 398} (1992) 407.
\bibitem{DNPT} D. Dicus, S. Nussinov, P.B. Pal, V.L. Teplitz,
{\it Phys. Lett.} {\bf B218} (1989) 84.
\bibitem{P} P.J.E. Peebles,  ``The Large Scale Structure of
the Universe'', (1980), Princeton University Press, Princeton (New Jersey).
\bibitem{W} S. Weinberg, {\it Ap. J.} {\bf 168} (1971) 175.
\bibitem{KTGR} E.W. Kolb, M.S. Turner, 
{\it Phys. Rev.} {\bf D 36} (1987) 2896.
S. Nussinov, M. Roncadelli, {\it Phys. Lett.} {\bf B 122} (1983) 157.
\bibitem{Pec} for a review of the physics of neutrinos, see
 R.D. Peccei, in {\it Proceedings of the Flavour
Symposium}, Beijing, China (1988),

 G.G. Raffelt, ``Stars as Laboratories for
Fundamental Physics'', (1996) University of Chicago Press, Chicago.
\bibitem{GR} G. Gelmini, M. Roncadelli, {\it Phys. Lett.} {\bf B 99}
(1981) 411.
\bibitem{Pal} P. Pal, {\it Phys. Lett.} {\bf B 205} (1988) 65.
\bibitem{CMP} Y. Chikashige, R.N. Mohapatra, R.D. Peccei,
{\it Phys. Lett.} {\bf B 98} (1981) 265.
\bibitem{sasaki} H. Kodama, M. Sasaki, {\it Prog. Theor. Phys.
Suppl.} {\bf 78} (1984) 1.
\bibitem{E} G. Efstathiou, in  {\it Physical Cosmology}.
J. Peacock, A.E. Heavens, A.T. Davies eds., (1990) Adam Hilger,
Bristol (England).
\bibitem{K+T} E.W. Kolb, M.S. Turner,  ``The Early Universe'', (1990)
Addison-Wesley.
\bibitem{K} K. Huang,  ``Statistical Mechanics'', (1987)
John Wiley, Second Edition, New York.
\bibitem{CE} S. Chapman, T.G. Cowling,  ``The Mathematical
Theory of Non-Uniform Gases'', (1970)
Cambridge University Press, Cambridge (England).
\bibitem{MAB} C.P. Ma, E. Bertschinger, {\it Astrophys. J.} {\bf 455} (1995) 7.
\bibitem{LL} L.D. Landau, E.M. Lifshitz,  ``Fluid Mechanics'', (1959)
Permagon Press. 
\bibitem{I+V} W. Isreal, J.N.Vardalas, {\it Let. Nuovo Cim.} {\bf IV} (1970)
887.
\bibitem{Mis} C.W. Misner, {\it Ap. J.} {\bf 151} (1968) 431.
\bibitem{T} L.H. Thomas, {\it Quart. J. Math.} (1930) 239.
\bibitem{MA}  C.P. Ma, E. Bertschinger, {\it Astrophys. J.} {\bf 455} (1994) 22.
\bibitem{GT} S. Tremaine, J.E. Gunn, {\it Phys. Rev. Lett.} {\bf 42} (1979) 407.
\bibitem{madsen1} J. Madsen, R.I. Epstein, {\it Astrophys. J.}
{\bf 282} (1984) 11.
\bibitem{madsen2} J. Madsen, {\it Phys. Rev. Lett.} {\bf 64} (1990)
2744.
\bibitem{gerhard} O.E. Gerhard, D.N. Spergel, {\it Astrophys. J.}
{\bf 389} (1992) L9.
\bibitem{lake} G. Lake, {\it Astrophys. J.} {\bf 356} (1990)  L43.
\bibitem{lucchin}  F. Lucchin {\it et. al.}, {\it Astrophys. J.} {\bf 459} (1996) 455.
\bibitem{Evrard} A.E. Evrard, F.J. Summers, M. Davis, {\it Astrophys. J.} {\bf 422} (1994) 11.
\end{thebibliography}
\end{document}